# Mechanism of Néel order switching in antiferromagnetic thin films revealed by magnetotransport and direct imaging

L. Baldrati[1]*, O. Gomonay[1], A. Ross[1,2], M. Filianina[1,2], R. Lebrun[1], R. Ramos[3], C. Leveille[1], F. Fuhrmann[1], T. R. Forrest[4], F. Maccherozzi[4], S. Valencia[5], F. Kronast[5], E. Saitoh[3,6,7,8,9], J. Sinova[1], M. Kläui[1,2]

[1]Institute of Physics, Johannes Gutenberg-University Mainz, 55128 Mainz, Germany

[2]Graduate School of Excellence Materials Science in Mainz, 55128 Mainz, Germany

[3]WPI-Advanced Institute for Materials Research, Tohoku University, Sendai 980-8577, Japan

[4]Diamond Light Source, Chilton, Didcot, Oxfordshire OX11 0DE, United Kingdom

[5]Helmholtz-Zentrum Berlin für Materialien und Energie, Albert-Einstein-Strasse 15, D-12489 Berlin, Germany

[6]Institute for Materials Research, Tohoku University, Sendai 980-8577, Japan

[7]Advanced Science Research Center, Japan Atomic Energy Agency, Tokai 319-1195, Japan

[8]Center for Spintronics Research Network, Tohoku University, Sendai 980-8577, Japan

[9]Department of Applied Physics, The University of Tokyo, Tokyo 113-8656, Japan

*Electronic Mail: lbaldrat@uni-mainz.de

## ABSTRACT

We probe the current-induced magnetic switching of insulating antiferromagnet/heavy metals systems, by electrical spin Hall magnetoresistance measurements and direct imaging, identifying a reversal occurring by domain wall (DW) motion. We observe switching of more than one third of the antiferromagnetic domains by the application of current pulses. Our data reveal two different magnetic switching mechanisms leading together to an efficient switching, namely the spin-current induced effective magnetic anisotropy variation and the action of the spin torque on the DWs.

# MANUSCRIPT

Electrical read-out and writing of the antiferromagnetic state is crucial to exploit the properties of antiferromagnets in future spintronic devices. Antiferromagnetic materials have the potential for ultrafast operation [1], with spin dynamics in the terahertz range, high packing density, due to the absence of stray magnetic fields, and an insensitivity to magnetic fields [2,3]. Furthermore, low-power operation is possible in antiferromagnetic insulators (AFM-Is) due to long spin diffusion lengths [4] and the theoretical prediction of superfluid spin transport [5].

Recently, the electrical reading of the Néel order (**n**) orientation in AFM-Is was demonstrated via spin Hall magnetoresistance (SMR) [6–10], a magnetoresistive effect depending on the mutual orientation of the magnetic order and an interfacial spin accumulation $\boldsymbol{\mu_s}$. However, one of the main challenges faced by AFM spintronics is the reliable electrical writing of the orientation of **n**. One possible approach exploits staggered Néel spin orbit torques [11], creating an effective field of opposite sign on each magnetic sublattice. However, these torques rely on special material requirements, which has limited their application to the conducting AFMs CuMnAs and $Mn_2Au$ [12–16]. Another approach would be to use the non-staggered, antidamping-like torque exerted by a spin accumulation at the interface of a heavy metal and an AFM-I. A charge current in the heavy metal layer can generate a transverse spin current via the spin Hall effect, creating antidamping-like torques in the antiferromagnet. The possibility of such switching was demonstrated in NiO(001)/Pt and Pt/NiO(111)/Pt [17,18], but the mechanisms are still debated. One of the possible mechanisms relies on spin-current induced domain wall (DW) motion [19], predicting that DWs with opposite chirality are driven in opposite directions, thus excluding the electrical signature of the switching when DWs with opposite chirality are equally probable. A second mechanism [18], based on the coherent rotation of **n**, predicts a current threshold ten times larger than that found experimentally. A third mechanism, based on field-like torques acting on uncompensated interfacial spins, requires perfectly flat interfaces [17]. Currently, none of these provides a consistent explanation of the effect.

In this work we realize reliable current-induced switching in epitaxial antiferromagnetic NiO/Pt bilayers. We show that the magnetic state of the NiO can be switched up to a thickness of at least 90 nm. By direct imaging of the current-induced switching, we single out the role of AFM DWs. Two switching mechanisms are identified to be involved, either breaking the degeneracy of **n** with respect to the spin accumulation $\boldsymbol{\mu_s}$ or not. We attribute the degeneracy-breaking mechanism to a ponderomotive force, created by the anti-damping like torque, which displaces the DWs and favors domains with **n** ⊥ $\boldsymbol{\mu_s}$. A second non degeneracy-breaking switching mechanism stems from the torque directly acting on the DWs, locally inducing switching in different directions (**n** ∥ $\boldsymbol{\mu_s}$, **n** ⊥ $\boldsymbol{\mu_s}$). These two mechanisms occur in AFMs with depinning fields of the DWs lower than the anisotropy fields, which is the case in NiO and most AFM-Is [9,10].

To study switching in AFM-Is, we grew epitaxial NiO(001)/Pt bilayers [20]. The magnetic properties were checked by the polarization-dependent absorption spectrum around the Ni $L_2$ edge (Fig. 1a), which shows x-ray magnetic linear dichroism (XMLD) and no circular dichroism (XMCD) [10,21,22], a signature of antiferromagnetic ordering. We read electrically the orientation of **n** by the SMR, since the transverse resistance of a heavy metal/AFM-I bilayer depends on the product $n_x*n_y$ [10]. To apply current pulses and measure the SMR, micrometric Hall cross devices were lithographically patterned and etched by Ar ions. For the SMR measurements, we applied a probing current density $j \sim 10^9$ A m$^{-2}$ and the relative transverse resistance variation was calculated as $\frac{\Delta R_{transv}}{\bar{R}} = \frac{V(I^+)-V(I^-)}{\bar{R}I}$, where $\bar{R}$ is the average longitudinal resistance, $V$ is the transverse voltage (Fig. 1b) and $I$ is the current, whose sign is reversed to eliminate thermal effect contributions. We applied current pulses and performed the measurements 10 seconds later, to probe equilibrium conditions.

The switching characteristics of an 8 μm wide Hall cross device on a MgO(001)//NiO(001)(5 nm)/Pt(2 nm) sample, obtained by changing the direction of the 1-ms long current pulses by 90° every five pulses is shown in Fig. 1c (a linear background was subtracted [20]). At 13 mA ($j = 8.1 \times 10^{11}$ A m$^{-2}$), the normalized transverse resistivity variation increases (decreases) after the application of current pulses along a direction at +45° (-45°) with respect to the measurement current direction. The first +45° pulse induces a "step-like" increase of the transverse resistance signal. The signal amplitude increases slightly with the following 4 pulses of the same orientation, with a tendency to saturate. At 15 mA +45° ($j = 9.4 \times 10^{11}$ A m$^{-2}$), the transverse resistance again increases abruptly as for smaller currents. However the signal decreases after the following pulses, implying a reversed sign of the switching and thus indicating the presence of at least two competing mechanisms contributing to the measured electrical signal. At even higher current densities only a "triangular-like" behavior is seen. In [20] we show that the "triangular-like" behavior at high currents is a thermal effect related to the Pt, observed also in MgO/Pt and in NiO/Pt with Pt grown ex-situ. The transverse resistance variation possibly stems from the current-induced annealing of the Pt deposited at room temperature, that locally changes the resistivity and yields different current paths in the system. On the other hand, we observed the "step-like" switching only in NiO/Pt with Pt grown in-situ, suggesting that this is related to the spin transport across the NiO/Pt interface and thus to the SMR probing the magnetic order in the NiO. While the switching depends on the pulse current orientation, it does not significantly depend on the polarity. The sign of the switching is consistent with the read-out by spin Hall magnetoresistance of a final state $\mathbf{n} \parallel \mathbf{j}$ [7,10], implying that the degeneracy between the $\mathbf{n} \perp \boldsymbol{\mu}_s$ and $\mathbf{n} \parallel \boldsymbol{\mu}_s$ configurations is broken. In Ref. [18], this switching was attributed to a spin-current induced antidamping-like torque acting in strained biaxial NiO(001), according to a macrospin model. However, the multi-level final state of the switching in contrast suggests that the switching comprises the redistribution of antiferromagnetic domains.

To develop a theory consistent with the experimental results, we first consider mechanisms based on the motion of AFM DWs. We start by the spin-current-induced dynamics of a simple antiferromagnetic texture comprising of two regions with a homogeneous direction of the Néel order, the domains A and B. These are separated by a DW, as shown schematically in Fig. 2. The orientation of $\mathbf{n}$ in two contiguous NiO domains can vary by different angles, due to the complex anisotropy of the material. We here consider 90° domains that are instructive to explain our model but the physical mechanisms is not limited to this situation. The translational motion of the DW has the lowest activation energy (zero in the absence of pinning) among all possible types of magnetic excitations and can be considered as the main mechanism of spin-current induced dynamics in AFMs with non-zero anisotropy like NiO. [9,10] In this case, the DW dynamics follows the equation of a point mass with momentum $\mathbf{P}$ [23]: $\frac{d\mathbf{P}}{dt} = -\gamma_d \mathbf{P} + \mathbf{F}_{\text{curr}} + \mathbf{F}_{\text{pin}}$, where $\gamma_d$ is the effective damping, $\mathbf{F}_{\text{pin}}$ is a pinning force, and $\mathbf{F}_{\text{curr}}$ is the force induced by the current, which is comprised of two components, as described below.

A charge current with a density $\mathbf{j}$ flowing in the Pt layer generates a damping-like spin-orbit torque (SOT) $\mathbf{T}_{\text{curr}} = \hbar\varepsilon\theta_H \mathbf{n} \times (\mathbf{j} \times \hat{\mathbf{z}}) \times \mathbf{n}/(2ed_{AF}M_s^2)$, acting on $\mathbf{n}$. Here $\hbar$ is the Planck constant, $d_{AF}$ is the thickness of the active layer of antiferromagnet, $0<\varepsilon \leq 1$ is the spin-polarization efficiency, $\theta_H$ is the spin Hall angle, $e$ is the electron charge, $M_s = |\mathbf{n}|$. In a homogeneous state, this torque competes with that $\mathbf{T}_{\text{an}} = \mathbf{n} \times \mathbf{H}_{\text{an}}$ created by the magnetic anisotropy field $\mathbf{H}_{\text{an}}$, and can rotate $\mathbf{n}$ from an easy axis towards a new equilibrium direction $\mathbf{n} + \Delta\mathbf{n}$ (Fig. 2a,b). The virtual work produced by the SOT in such static rotation is associated with the potential energy density $U_{\text{curr}} = \hbar\varepsilon\theta_H(\hat{\mathbf{z}} \times \mathbf{j}) \cdot (\mathbf{n} \times \Delta\mathbf{n})/(2ed_{AF}M_s^2)$, i.e. the spin current acts like an additional magnetic anisotropy term which depends on **n**: $U_{\text{ma}} \to U \equiv U_{\text{ma}} + U_{\text{curr}}$ [20]. The resulting energy imbalance between the two domains entails a force which drives the DW into the energetically unfavorable domain. We call this force due to its nature the ponderomotive force $F_{\text{pond}} = U(\mathbf{n}_A) - U(\mathbf{n}_B) \propto (\boldsymbol{j} \cdot \mathbf{n}_B)^2 - (\boldsymbol{j} \cdot \mathbf{n}_A)^2$. In a multidomain sample, **F**$_{\text{pond}}$ breaks the degeneracy of the domains with different **n** and thus induces switching toward a state with **n** ⊥ $\boldsymbol{\mu_s}$, as we observe here experimentally. Note that the thermally activated processes in our theory can be modelled as a temperature dependent pinning force **F**$_{\text{pin}}$, which decreases with increasing temperature.

To further investigate the role that the antiferromagnetic domains and DWs play in the switching mechanism, we performed XMLD-photoemission electron microscopy (PEEM) imaging of the NiO domains in NiO(001) samples [22], grown at the same time as the ones for electrical measurements, while applying *in-situ* current pulses. The imaging was performed using a two energy mode at the Ni $L_2$ double peak [21], using linearly polarized x-rays with the electric field out of the plane of the sample (Fig. 1a and Ref. [20]), yielding sensitivity to components of **n** parallel/orthogonal.

We show in Fig. 3a-i the domain structure of a MgO//NiO(10 nm)/Pt(2 nm) sample, before and after the application of pulses across two orthogonal arms of a Hall cross, as measured at the SPEEM endstation at Helmholtz-Zentrum Berlin [24]. We applied sequences of 5 pulses 1 ms long with currents of +28 mA (Fig. 3a-c, $j = 1.4 \times 10^{12}$ A m$^{-2}$), -28 mA (Fig. 3d-f) and +31 mA (Fig. 3g-i, $j = 1.5 \times 10^{12}$ A m$^{-2}$). We first note that, after the application of the +28 mA pulse train, the contrast changes in approximately one third of the area, towards more white contrast (Fig. 3a-c). Given the formula used to calculate the contrast [20], the final state has increased areas with **n** out of the plane of the sample (parallel to the x-ray polarization), consistent with our model predicting a final state with **n** ⊥ $\boldsymbol{\mu_s}$. Moreover, a large domain area goes instead toward more black (in-plane). We cannot resolve in-plane components of **n** with this measurement configuration, but there is an in-plane direction with **n** ⊥ $\boldsymbol{\mu_s}$. Pulses with current lower than 28 mA did not change the domain structure significantly. One can see that some domains shrink after the pulse train, while other domain walls do not move, as described in our model by the space dependent pinning force. Reversing the current sign and applying 5 additional current pulses (Fig. 3d-f) yields again more white domains, consistently with the independence on the pulse current polarity and the tendency to saturate found in electrical measurements. Finally, at even larger current density (Fig. 3g-i) we observe additional switching toward more out-of-plane domains, showing that the switching is deterministic and increases with increasing current density, in line with our model.

In addition to this unidirectional deterministic switching, further switching mechanisms have been predicted that change the domain structure but keep the average distribution of **n** constant, so they cannot be detected by electrical means. To check if this is the case, we imaged the domain structure of a MgO//NiO(25 nm)/Pt(2 nm) sample with in-plane x-ray polarization, before and after the application of 1000 current pulses 10 µs long with a lower current density of $7.5 \times 10^{11}$ A m$^{-2}$, where no significant switching is detected electrically. Such a switching event is shown in Fig. 4a-c, together with the difference image, as measured at the beamline I06 of Diamond light source. One can see sub-µm sized antiferromagnetic domains switching after the application of the pulses. In particular, we observe switching having in-plane components in both directions (**n** ∥ $\boldsymbol{\mu_s}$, **n** ⊥ $\boldsymbol{\mu_s}$) for a single pulse direction. To check for pure thermal effects, we imaged previously the domain structure as a function of temperature and did not observe pure thermal switching of the antiferromagnetic domains [10], implying that the switching observed here is current-induced due to generated torques. The switching mechanism observed here, not breaking the degeneracy between the (**n** ∥ $\boldsymbol{\mu_s}$, **n** ⊥ $\boldsymbol{\mu_s}$) states, is not explained by the antidamping-torque theory [18], and it is not consistent with the symmetry of the ponderomotive force, thus calling for an additional theoretical explanation. For this second switching mechanism, in analogy to ferromagnets [25], we identify the SOT acting in inhomogeneous regions of the antiferromagnetic texture and inducing a coherent rotation of the spatially distributed **n**, i.e. leading to translational DW motion (Fig. 3c) induced by a force $\mathbf{F}_{\text{DW}}$. This force [19], (see derivation in [20])

$$\mathbf{F}_{\text{DW}} = \frac{\hbar \varepsilon \theta_H}{2 e d_{AF} M_s^2} \int (\hat{\mathbf{z}} \times \boldsymbol{j}) \cdot (\mathbf{n} \times \nabla \mathbf{n}) dx, \qquad (1)$$

originates from the current-induced rotation of **n** within the DW,[1] is linear with the current and its direction depends only on the chirality of the DW (**n** × ∇**n**) and not on $\mathbf{n}_A$ and $\mathbf{n}_B$ inside the domains. $\mathbf{F}_{\text{DW}}$, though able to locally induce fast motion of the DWs, does not globally break the degeneracy of the domains between the configurations (**n** ∥ $\boldsymbol{\mu_s}$, **n** ⊥ $\boldsymbol{\mu_s}$) once the DWs with opposite chirality are equiprobable, resulting in no electrical response. This is expected in NiO, due to absence of interactions such as the Dzyaloshinskii-Moriya, breaking the chiral degeneracy.

---

[1] The DW dynamics induced by this force was considered in Ref. [19], but not the general expression considered here.

Overall, both mechanisms identified from the combination of electrical measurements above the threshold and the imaging below the electrical threshold contribute to the switching. The current-induced force is thus $\mathbf{F}_{curr} = \mathbf{F}_{pond} + \mathbf{F}_{DW}$. The resulting $\mathbf{F}_{curr}$ acting on the DWs depends on the orientation of the current (spin polarization) with respect to the easy plane. If the current is almost parallel to the easy plane, $|\mathbf{F}_{DW}| \sim |\mathbf{F}_{pond}|$, the motion of the DWs into energetically favorable domains can be partially or fully blocked for one DW chirality, depending on the value of $F_{pin}$ (Fig. 3d). Note that $\mathbf{F}_{DW}$, not breaking the degeneracy ($\mathbf{n} \parallel \boldsymbol{\mu}_s, \mathbf{n} \perp \boldsymbol{\mu}_s$), is not expected to lead to an electrical signal. This is distinctly different from the mechanism proposed in Ref. [17] for Pt/NiO(111)/Pt trilayers, based on the field-like torque acting on the uncompensated spins at the interface, which are unlikely to form in our non-perfectly flat devices.

In the case of a pronounced angle between the easy plane and the film plane, as we have in NiO(001), the system exhibits $|\mathbf{F}_{DW}| > |\mathbf{F}_{pond}|$ at low current densities, and local switching in both directions (A to B or B to A) is possible. This is consistent with the direct observation by XMLD-PEEM of switching into different final states ($\mathbf{n} \parallel \boldsymbol{\mu}_s$ and $\mathbf{n} \perp \boldsymbol{\mu}_s$) at $7.5 \times 10^{11}$ A m$^{-2}$ in NiO(001). However, at higher current densities $\mathbf{F}_{pond} \propto I^2$ prevails over $\mathbf{F}_{DW} \propto I$, as shown in Fig. 2d, and drives the deterministic switching as we see in Fig. 3. Antiferromagnetic DW motion induced by thermal gradients might also aid the switching process and lead to additional final states ($\mathbf{n} \parallel \boldsymbol{\mu}_s, \mathbf{n} \perp \boldsymbol{\mu}_s$) [26,27].

We finally compare the "step-like" switching in MgO(001)//NiO(d)/Pt(2 nm) samples, where d = 5, 90 nm (see [20]). The switching amplitude is larger and the maximum of the switching occurs at lower current densities for d = 5 nm. The easier switching in the thinner NiO layer can be explained by the reduced volume to be switched and by the smaller domains we observe in thicker NiO (see [20]), which indicate a higher density of pinning defects. Moreover, we can switch the NiO(001) up to a thickness of 90 nm, far beyond the highest spin-orbit torque switchable thickness reported in ferromagnetic Tm$_3$Fe$_5$O$_{12}$(8 nm)/Pt [28]. This allows us to speculate that the switching in NiO occurs in the interfacial region close to the Pt layer, i.e. the formation of surface domains in antiferromagnets is easier than in ferromagnetic systems, due to stronger destressing effects and the presence of dislocations at the surface [29]. These surface domains can be as small as the effective spin diffusion length of the NiO [30], of the order of few nm [31], i.e. the depths probed by the transverse SMR and XMLD-PEEM measurements.

To conclude, we demonstrated current-induced switching of the Néel order in the NiO/Pt system, revealing the origin of the switching. The switching comprises the redistribution of antiferromagnetic domains via domain wall motion, as probed both by electrical measurements and direct magnetic imaging, and occurs via two different mechanisms: one mechanism breaks the degeneracy of the domains ($\mathbf{n} \parallel \boldsymbol{\mu}_s, \mathbf{n} \perp \boldsymbol{\mu}_s$) and stems from the action of the spin current, which modifies the effective magnetic anisotropy, determining a ponderomotive force on the domain walls that leads to the switching detected by the electrical measurements above a threshold. The second mechanism stems from the direct action of the antidamping spin torque on the domain walls and does not break the degeneracy ($\mathbf{n} \parallel \boldsymbol{\mu}_s, \mathbf{n} \perp \boldsymbol{\mu}_s$) if domain walls with different chirality are equally probable, as identified from imaging below the electrical threshold. Our model has the potential to explain switching in antiferromagnetic systems in which antidamping-like spin torques can be created, thus paving the way to the control of the switching necessary to enable future applications of AFMs in devices.

*Note added*: During the preparation of this revised manuscript, we became aware of related work by Gray et al. [32].

**Acknowledgements**

The authors thank J. Cramer, M. Asa, J. Henrizi, A. Dion, T. Reimer for skillful technical assistance. We acknowledge useful scientific discussion with M. Jourdan. O.G. and J.S. acknowledge the support from the Humboldt Foundation, the ERC Synergy Grant SC2 (No. 610115), the EU FET Open RIA Grant no. 766566, the DFG (project SHARP 397322108), from the Ministry of Education of the Czech Republic Grant No. LM2015087 and LNSM-LNSpin. L.B., R.L., A.R., M.F. and M.K. acknowledge support from the Graduate School of Excellence Materials Science in Mainz (MAINZ) DFG 266, the DAAD (Spintronics network, Project No. 57334897) and all groups from Mainz acknowledge SFB TRR 173 Spin+X. L.B and R.L. acknowledge the European Union's Horizon 2020 research and innovation program under the Marie Skłodowska-Curie grant agreements ARTES number 793159 and FAST number 752195. We acknowledge Diamond Light Source for time on beamline I06 under proposals SI18850 and SI20698. The research leading to this result has been supported by the project CALIPSOplus under the Grant Agreement 730872 from the EU Framework Programme for Research and Innovation HORIZON 2020. This work was also supported by ERATO "Spin Quantum Rectification Project" (Grant No. JPMJER1402) and the Grant-in-Aid for Scientific Research on Innovative Area, "Nano Spin Conversion Science" (Grant No. JP26103005) from JSPS KAKENHI, Japan.

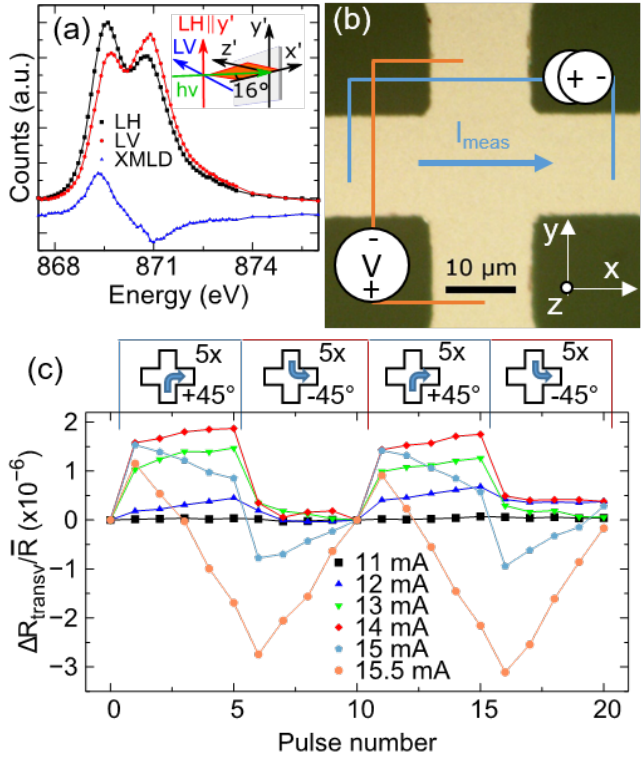

Fig. 1: (a) X-ray absorption spectrum at the Ni $L_2$ edge for linearly vertical (LV) and horizontal (LH) polarized light of MgO(001)//NiO(25 nm)/Pt(2). (b) Optical micrograph of a device and contact scheme used for the transverse resistance measurements. (c) Electrical switching of the transverse resistance in a MgO(001)//NiO(5 nm)/Pt(2 nm) sample. The pulse pathway is changed every 5 pulses as indicated.

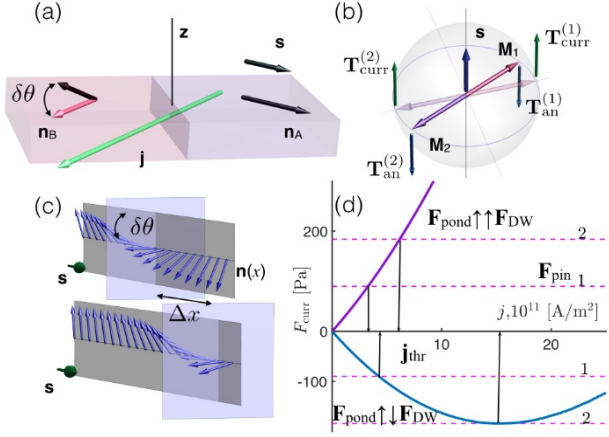

Fig. 2: (a) A current with density $\mathbf{j} \perp \mathbf{n}_A$ injects a spin current with polarization $\mathbf{s} \propto (\mathbf{j} \times \hat{\mathbf{z}}) \parallel \mathbf{n}_A$, creating a torque $\mathbf{T}_{\text{curr}}$. The torque rotates $\mathbf{n}_B$, initially $\parallel \mathbf{j}$, by an angle $\delta\theta$, but does not affect $\mathbf{n}_A$. (b) Due to the SOTs, $\mathbf{M}_{1,2}$ rotates from the easy axis (semitransparent arrows) toward the new equilibrium state (opaque arrows) where $\mathbf{T}_{\text{curr}}^{(1,2)}$ are compensated by the anisotropy torques $\mathbf{T}_{\text{an}}^{(1,2)}$. (c) The SOT-induced translation of the DW by a distance $\Delta x$ is equivalent to the rotation of $\mathbf{n}$ inside the DW region by an angle $\delta\theta$. (d) Current dependence of $\mathbf{F}_{\text{curr}}$, when $\mathbf{s}$ is almost parallel to easy plane (deflection 5°). The force pushes the DWs toward the unfavorite domain ($\mathbf{F}_{\text{pond}} \uparrow\uparrow \mathbf{F}_{\text{DW}}$), but for low current density and low pinning force (dashed line 1) the DW is pushed toward the favorite one ($\mathbf{F}_{\text{pond}} \uparrow\downarrow \mathbf{F}_{\text{DW}}$). A large pinning force (dashed line 2) blocks the DW motion.

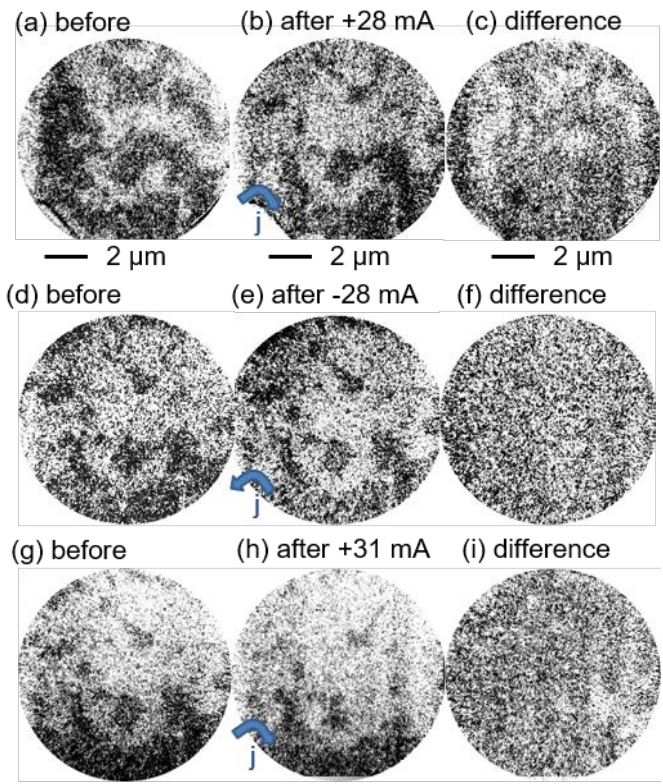

Fig. 3: Switching of antiferromagnetic domains in MgO//NiO(10)/Pt(2), imaged with out-of-plane x-ray polarization. Three sequences of images before and after 5 pulses 1 ms long are shown together with the difference image. The direction of the current density j is shown in blue in panels b, e, h.

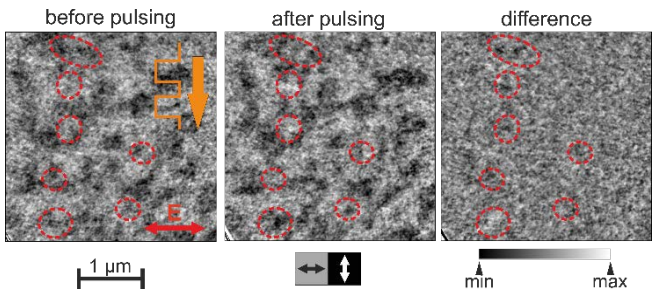

Fig. 4: Switching of antiferromagnetic domains in MgO//NiO(25)/Pt(2) imaged with in-plane x-ray polarization. The NiO domain structure is shown (a) before and (b) after the application of 1000 pulses 10 μs long, with a current density of $7.5 \times 10^{11}$ A m$^{-2}$. (c) Difference between the images in panels a,b. Switching areas, showing different final states (**n** ∥ **j**, **n** ⊥ **j**) are encircled.

# Mechanism of Néel order switching in antiferromagnetic thin films revealed by magnetotransport and direct imaging – supplementary information

L. Baldrati[1]*, O. Gomonay[1], A. Ross[1,2], M. Filianina[1,2], R. Lebrun[1], R. Ramos[3], C. Leveille[1], F. Fuhrmann[1], T. R. Forrest[4], F. Maccherozzi[4], S. Valencia[5], F. Kronast[5], E. Saitoh[3,6,7,8,9], J. Sinova[1], M. Kläui[1,2]

[1]*Institute of Physics, Johannes Gutenberg-University Mainz, 55128 Mainz, Germany*

[2]*Graduate School of Excellence Materials Science in Mainz, 55128 Mainz, Germany*

[3]*WPI-Advanced Institute for Materials Research, Tohoku University, Sendai 980-8577, Japan*

[4]*Diamond Light Source, Chilton, Didcot, Oxfordshire OX11 0DE, United Kingdom*

[5]*Helmholtz-Zentrum Berlin für Materialien und Energie, Albert-Einstein-Strasse 15, D-12489 Berlin, Germany*

[6]*Institute for Materials Research, Tohoku University, Sendai 980-8577, Japan*

[7]*Advanced Science Research Center, Japan Atomic Energy Agency, Tokai 319-1195, Japan*

[8]*Center for Spintronics Research Network, Tohoku University, Sendai 980-8577, Japan*

[9]*Department of Applied Physics, The University of Tokyo, Tokyo 113-8656, Japan*

*\*Electronic Mail: lbaldrat@uni-mainz.de*

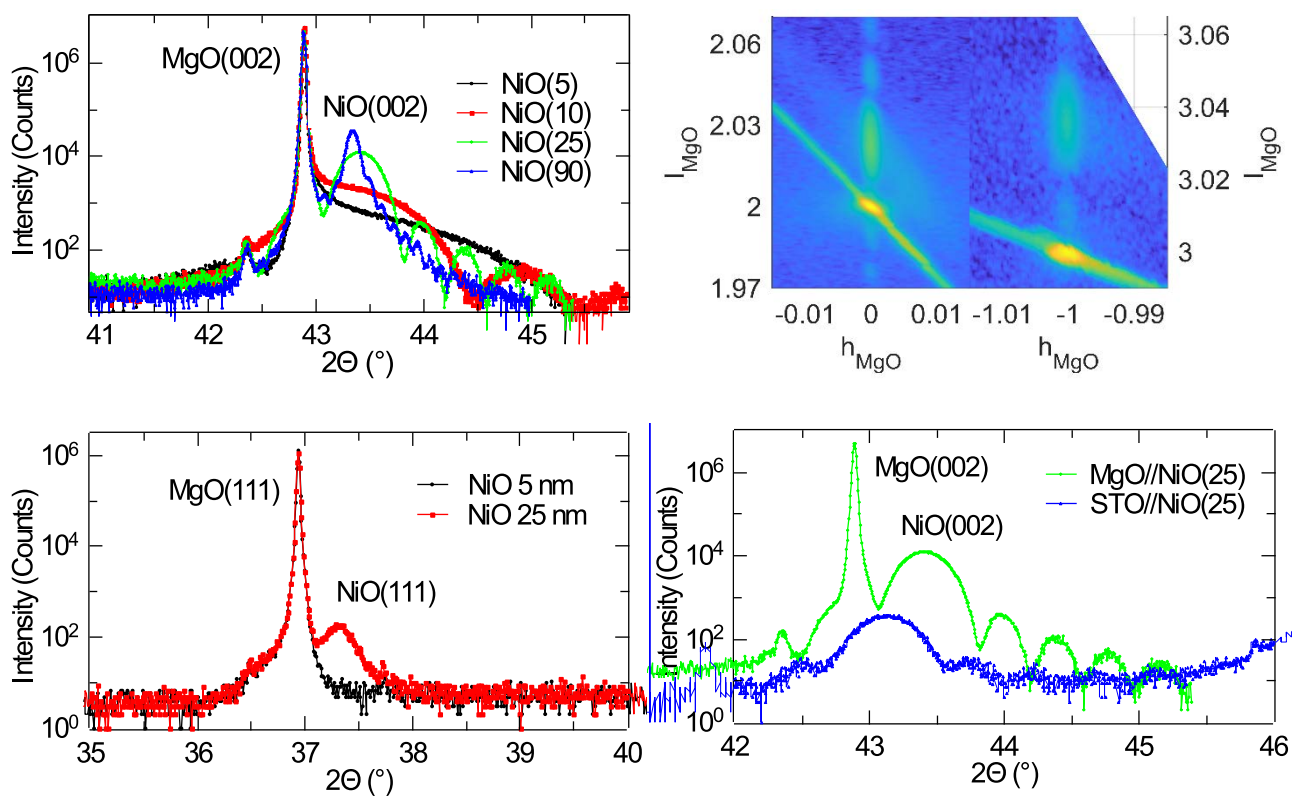

*Figure S1: (a) 2θ-ω x-ray diffraction scans of NiO(d)/Pt(2) thin films grown on MgO(001) substrates. (b) symmetric reciprocal space map (RSM) measurements around the MgO(002) peak. (c) antisymmetric RSM*

**Growth of epitaxial NiO thin films and structural characterization**

We grew high quality epitaxial NiO(001) thin films on 0.5 mm thick MgO(001) 20x20 mm² substrates and SrTiO$_3$(001) (STO) 14x14 mm² substrates by reactive sputtering of a Ni target in mixed Ar (flow 15 sccm) and O$_2$ (flow 1.5 sccm) atmosphere, at 430 °C. NiO(111)/Pt bilayers were grown in the same conditions on MgO(111) substrates. The Pt layer was deposited at room temperature, after cooling down the samples in vacuum. Symmetric x-ray diffraction 2θ-ω scans of NiO(d)/Pt(2) thin films, where d = 5, 10, 25, 50, 90 nm, grown on MgO(001) substrates were acquired using a Bruker D8 Discover high resolution diffractometer, using Cu K$_\alpha$ radiation, around the substrate MgO(002) peak at 2θ = 42.91° (substrate lattice constant 4.212 Å). The results, shown in Fig. S1a, comprise strong peaks and Laue oscillations, indicating the high crystalline epitaxial quality of the NiO layers. The increasing out-of-plane lattice constant as a function of the NiO layer thickness, suggests a high degree of residual strain in the thinner (d = 5, 10 nm) NiO layers, that is relaxed at higher NiO thickness. The position of the NiO(002) peak of the 90-nm thick NiO sample indicates an out-of-plane lattice constant of 4.172 ± 0.001 Å, similar to the bulk NiO lattice constant value of 4.176 Å, [1] while smaller lattice constants are seen in samples comprising thinner NiO samples, indicating the expected presence of in-plane tensile strain from the MgO(001) substrates. By analyzing the NiO(002) peak widths with the Scherrer formula τ=kλ/(FWHM*θ), one can obtain the mean size of the crystallites τ. k is the shape factor of the NiO crystal (0.9), λ is the wavelength of the x-rays, FWHM is the full width at half maximum of the XRD peak and θ(=2θ/2) is the diffraction angle. The value of τ (for τ < thickness) can be related to the density of stacking defects along the film thickness. We found for the 90 nm thick NiO film a value τ=78 nm (87% of the thickness) and for the 25 nm thick NiO τ=23 nm (92% of the thickness). While the crystalline quality is thus good in both cases, the lower relative value of τ for the 90 nm thick NiO film suggests that the thicker films have a higher density of stacking defects. To additionally probe the crystalline structure of the NiO thin films we performed symmetric and antisymmetric reciprocal space mapping measurements on a MgO(001)//NiO(25 nm)/Pt(2 nm) sample, at the MgO(002) and (113) diffraction peaks. The vertical alignment along the same h-value of the MgO and NiO peaks, as shown in the measurements in Fig. S1b,c, indicates a polymorphic growth of the NiO layer, with the same in-plane lattice constant as the MgO(001) substrate, while the vertically aligned minor peaks are Laue oscillations, further evidencing the crystallinity of the samples. We also show in Fig. S1d the XRD curves of NiO(111)/Pt bilayers, where the NiO(111) peak can be seen at 37.34° in the sample comprising a 25 nm thick NiO layer. Finally, we compare the XRD patterns of NiO(001) layers 25 nm thick grown on MgO(001) and STO(001). One can see that the NiO layer grown on MgO(001) presents a peak corresponding to a lattice constant of 4.166 Å, while the NiO layer grown on STO(001) (substrate lattice constant 3.905 Å) [2] presents an out-of-plane lattice constant equal to 4.192 Å, higher than the bulk value, as it is expected due to the compressive in-plane strain experienced by the NiO layer.

**Switching data and background subtraction**

The raw switching data of a MgO//NiO(10)/Pt(2) sample, where a Hall cross 11 µm wide was patterned, are shown in Fig. S2a, using the same pulse configurations as in Fig. 2 of the main text, with different current flow directions (±45°), alternated every 5 pulses. A measurement current of density ~10$^9$ A/m² was applied by a Keithley 2400, while acquiring the transverse voltage by a Keithley 2182A nanovoltmeter, as shown in Fig. 1b of the main text. Current pulses were applied by a Keithley 6221 AC/DC current source both for the electrical measurements (in ambient air at room temperature) and the imaging. The electrical measurements were performed in ambient air at room temperature, measuring the transverse resistance 10 s after the pulse. As one can see, both a symmetric (not depending on the pulse orientation) and antisymmetric switching components (depending on the pulse orientation) exist. The same data in Fig. S2a, after subtraction of the symmetric component of the switching (subtraction of a straight line to each set of 20 pulses), are presented in Fig. S2b. The origin of the antisymmetric switching component is extensively described in the main text. The origin of the symmetric switching component is at the moment not clear, but possibly depends on the geometry of the device and pulsing configuration used, where the current pulses sent at +45° and -45° have one arm in common, so that the current orientation in the region close to this arm is similar for the two pulse types. Moreover, from these data, one can see that the antisymmetric switching amplitude decreases versus the number of applied pulses, at a fixed current density, i.e. the device efficiency decreases. We attribute this effect to the switching at the corners of the cross, as was discussed for the switching of $Mn_2Au$, [3] since the current flows always in the same direction close to the corners, even when one changes the configuration between the +45° and -45° pulses. Due to the type of devices used, it is not possible to switch those regions back to the opposite configuration.

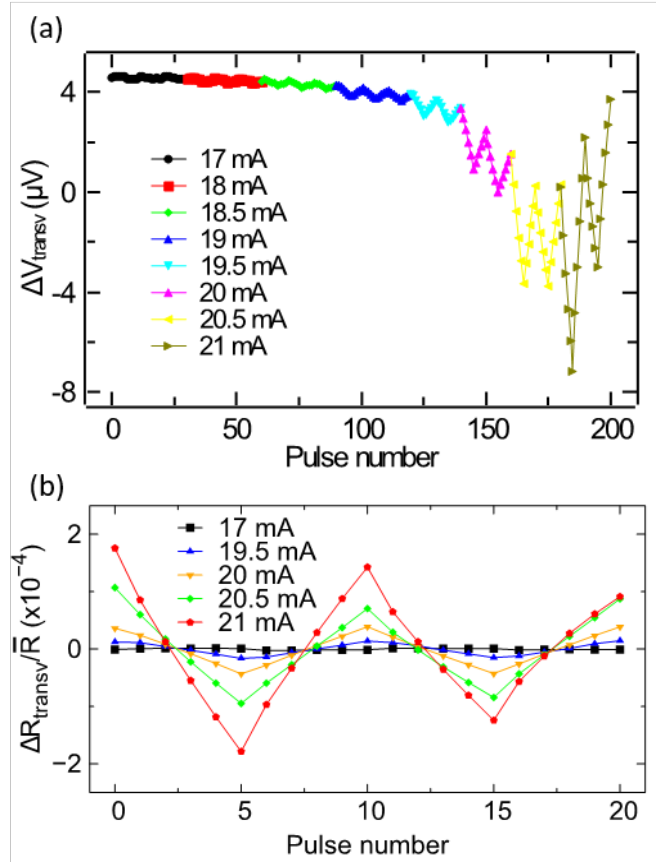

*Figure S2: (a) Raw switching data expressed in terms of transverse voltage, changing the pulse orientation between +45° and -45° configuration*

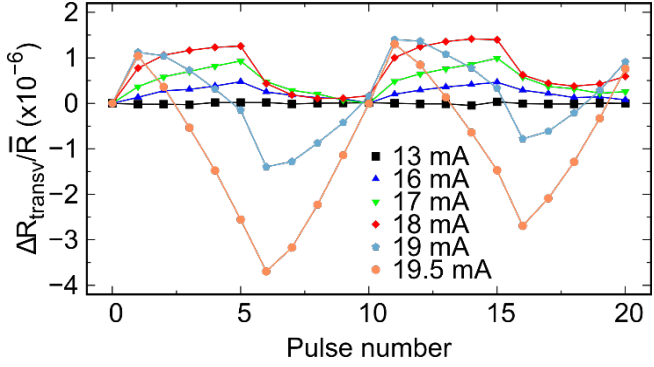

*Figure S3: Electrical switching of the transverse resistance in a MgO(001)//NiO(90 nm)/Pt(2 nm)*

### Switching NiO(90 nm)/Pt

We measured the current dependence of the switching in a MgO(001)/NiO(90 nm)/Pt(2 nm), using the same sequence as described in Fig. 1c of the main text for a Hall cross 11.5 micron wide in the virgin state, as shown in Fig. S3. One can see that the measurable switching amplitude is lower and pulse current at maximum larger compared to the 5 nm sample presented in the main text, indicating a more difficult switching in samples with thicker NiO layer.

### Switching in MgO/Pt and NiO/Pt with Pt grown ex-situ

To check the dependence of the switching on the magnetic properties and interface quality, we covered half of a MgO(001)//NiO(30 nm) sample with a thin 2-nm MgO layer and subsequently deposited Pt ex-situ on top of the whole sample, so that in 50% of the devices the Pt layer was not in direct contact with the antiferromagnetic layer and in the other 50% the Pt layer was in contact to NiO, but with poor interface quality due to the ex-situ growth. In Fig. S4 we show the switching behavior of two nominally identical devices in the half of the sample covered by MgO (Fig. S4a) or not (Fig. S4b). In none of these devices we observed the low current step-like switching seen in Fig. 1c of the main text, while we observe the triangular-like switching like in the samples with Pt grown in-situ at higher current densities. This clearly indicates that the triangular-like switching is related to the Pt layer, while the low current behavior is likely related to the magnetic properties of the NiO, since it depends on the higher quality of the NiO/Pt interface that is obtained by in-situ deposition. The resistance of the Pt grown on MgO is larger (8.5 kOhm) compared to the one of the Pt grown on the NiO (2.8 kOhm), which correlates with the different threshold in the two cases. This is possibly due to different Pt thicknesses arising from the different sticking coefficients during the deposition (lower threshold corresponds to higher resistance). The data were acquired in devices after the application of 1000 pulses at a current density in the triangular-like switching regime, but the behavior is similar to the one observed in the virgin state, considering that switching is seen at a lower current threshold in virgin samples.

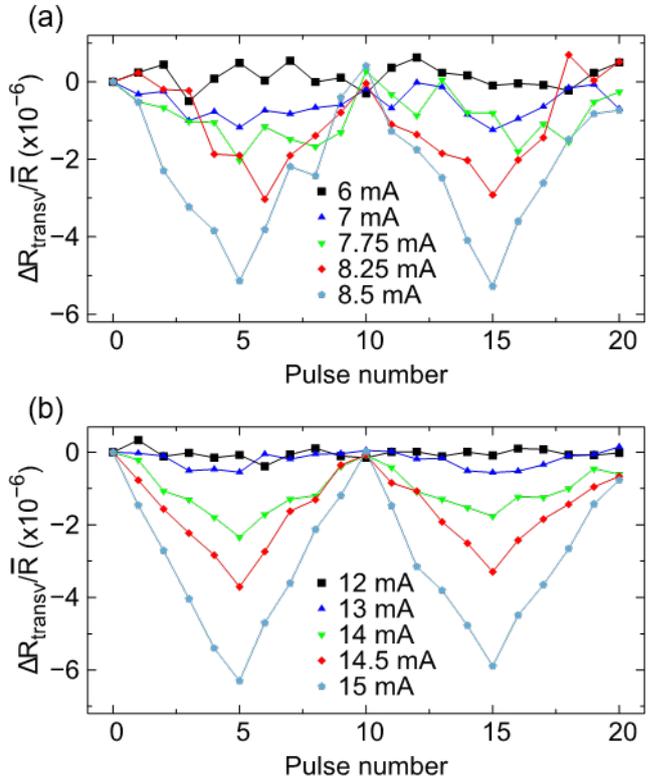

*Figure S4: (a) switching in the MgO//NiO(30)/MgO(2)/Pt(2) half of the sample. (b)*

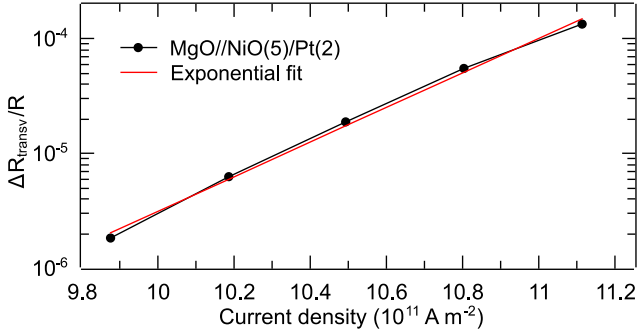

*Fig. S5: Switching signal amplitude (absolute value) as a function of the current density above*

**Exponential increase of the triangular-like signal amplitude around threshold**

To show clearly the dependence of the amplitude of the Pt-related triangular-like signal with respect to the current density around the threshold current density, we present in Fig. S5 the MgO//NiO(5)/Pt(2) nm data, in a log scale and together with the exponential best-fitting curve. The fitting function used is $\frac{\Delta R}{R} = a\, exp(\frac{I}{I_0})$, and we found fit parameters $a = 1.97 \pm 0.06$ and $I_0 = (2.9 \pm 0.1) \times 10^{10}$ A m$^{-2}$.

**Orientation dependence of the switching and equivalence of the "2 arms" and "4 arms" pulsing configurations**

The orientation dependence of the switching and the equivalence between the "2 arms" pulsing configuration used in this study and the previously reported "4 arms" configuration, [4] are shown in Fig. S6a-f, as measured in a Hall cross 10 µm wide patterned on a MgO//NiO(25)/Pt(2) sample. Pulses with current directions differing by 90° rotations with respect to the measurement current (±45°, ±135°) have been applied to the patterned Hall cross, and the different pulsing configurations are shown in Fig. S6c,f. We used a current equal to 28 mA (current density $9.9 \times 10^{11}$ A m$^{-2}$) in the "4 arms" configuration and equal to 19 mA (and $9.5 \times 10^{11}$ A m$^{-2}$) in the "2 arms" configuration. The current density was calculated by taking into account the increased cross sectional area by a $\sqrt{2}$ factor when the "4 arms" configuration is used (diagonal of the cross), with respect to the "2 arms" configuration. This is also reflected in the current density threshold in the two configurations, which we experimentally find to be the same only if we take into account the $\sqrt{2}$ factor. As one can see in Fig. S6, the switching sign and amplitude are predominantly defined by the orientation of the current, while the current polarity has no significant effect. Moreover, the pulsing configuration comprising 2 or 4 arms gives similar switching patterns. The slightly different amplitudes obtained by switching the current polarity are not easily comparable at this stage, due to the difficulty of defining a reproducible initial state, so that further study is required to fully understand if the current polarity plays an actually relevant role at all in the switching. Moreover, the two configurations seem equally efficient, if one considers that the higher transverse resistivity variation obtained in the 4 arms configurations uses a pulse energy which is approximately twice as large as the one used in the "2 arms" configuration. We also tested crosses rotated by 45° with respect to the substrate (001) direction and found consistent results.

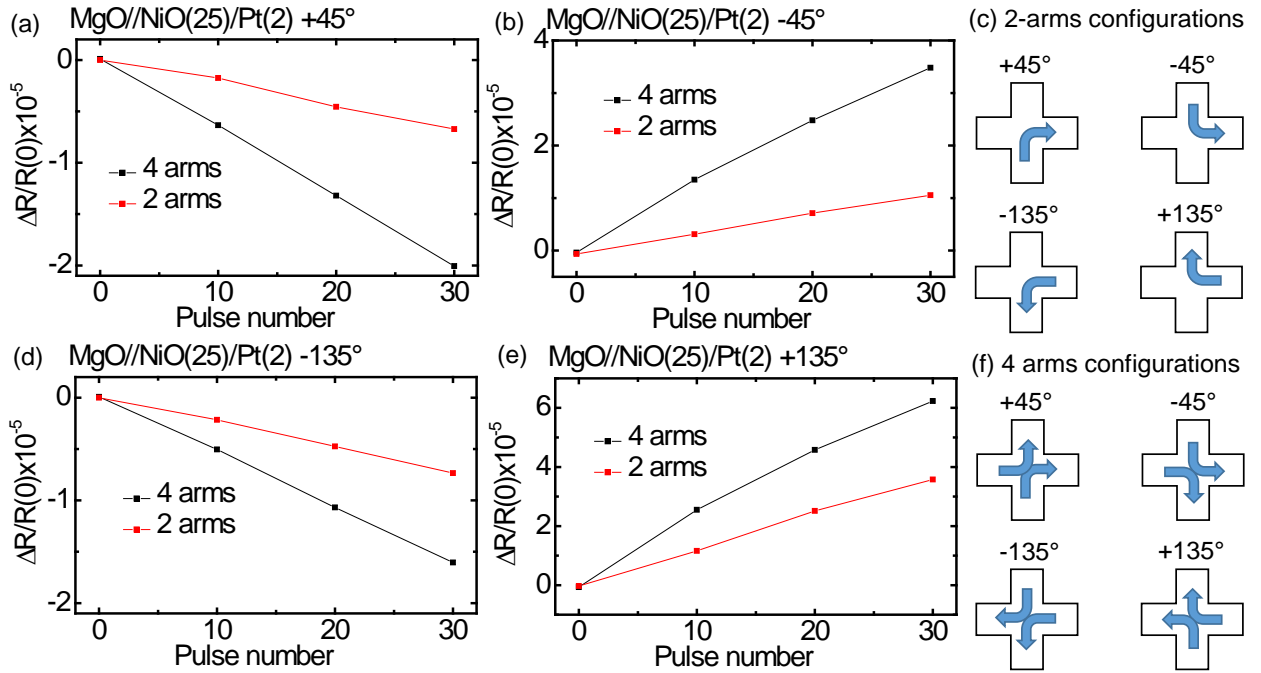

*Figure S6: Orientation and configuration dependence of the current induced switching of a 10 µm wide Hall cross patterned on top of a MgO//NiO(25)/Pt(2) sample. Switching at (a) +45° (b) -45° (d) -135° (e) +135°.*

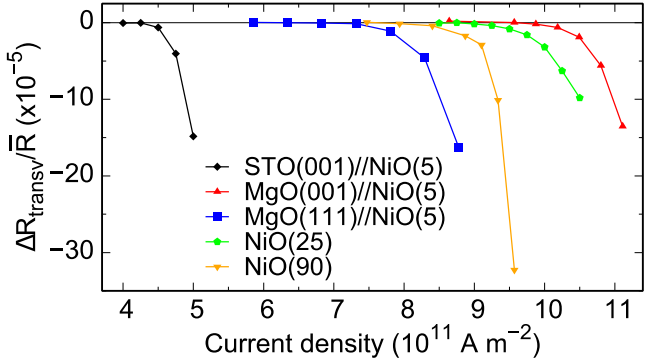 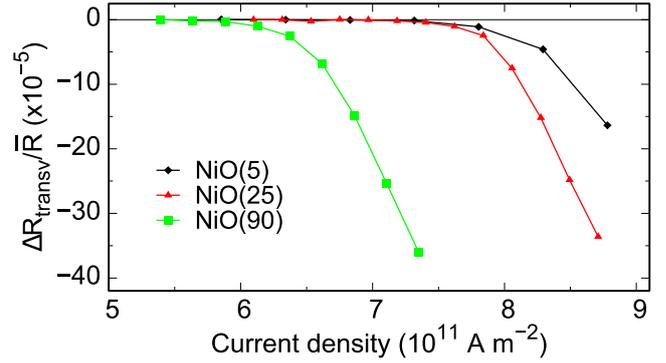

*Fig. S7: Current-induced triangular-like switching amplitude in STO(001)//NiO(001)/Pt, MgO(001)//NiO(001)/Pt and MgO(111)//NiO(111)/Pt devices as a function of*

*Figure S8: NiO(111) current density and thickness dependence of the triangular-like switching amplitude in NiO(d)/Pt samples, grown on*

**Thickness and substrate dependence**
To check if the Pt-related triangular-like switching is facilitated by the strain applied to the NiO [5], we grew additional NiO(001)/Pt bilayers on STO(001) (lattice mismatch -6.9%, compressive strain) and on MgO(001)

(mismatch +0.9%, tensile strain). In Fig. S7, we show the transverse SMR amplitude as a function of the pulse current density, after the application of four trains of 5 pulses 1 ms long with alternating current orientation. All transverse resistance curves are exponential above a current density threshold, which is half that of the MgO substrate for the STO substrate. Previously, it was proposed that the triangular-like switching was possible in biaxial STO(001)//NiO(001), but was prevented in MgO(001)//NiO(001) [5]. However, we show here that triangular-like switching on both substrates is possible, in contrast to what was found in Ref. [5]. To reveal the origin of the different threshold current density, we consider the higher thermal conductivity of MgO (57 W m$^{-1}$ K$^{-1}$) compared to STO (11 W m$^{-1}$ K$^{-1}$) [6]. In both cases the current density threshold corresponds to a temperature increase of 35±10 K, since pulses of equal current density lead to lower temperature variations in the NiO/Pt Hall crosses grown on MgO. Due to the larger heat dissipation, higher current densities are needed for switching on MgO, which could explain why no switching was found in Ref. [5]. Note that the exponential current dependence of the transverse resistance above threshold, shown above, indicates a thermally activated triangular-like switching process, as is the case for switching in conducting AFMs [3] and AFM/FM bilayers [7], and this is in contrast to ferromagnets, where the switching relies on the direct spin torque effect of the current [8].

Next, to probe if a different switching mechanism occurs in NiO(111), we consider switching of NiO(111)/Pt on MgO(111), as shown in Fig. S8. One can see that this system has an analogous behavior as the one of NiO(001) on MgO(001) and STO, with an intermediate current threshold, implying that all the three systems considered here present the same triangular-like switching. We additionally performed switching measurements for a series of MgO(111)/NiO(t)/Pt(2) nm samples, as shown in Fig. S8. One can see that the current threshold of the switching is consistently lower than the threshold observed in NiO(001) grown on MgO for the same thickness. Moreover, the current density threshold is minimum in the NiO(111) 90 nm thick sample and maximum in the 5 nm thick, as it occurred in NiO(001). In both cases, the lower current threshold can originate from a different quality of the Pt grown on thicker NiO layers, while the higher roughness of the NiO(111) and hence of the Pt might originate the lower threshold. Since the triangular-like switching is not related to the magnetic properties, we conclude that antiferromagnetic switching can be better seen electrically in thin NiO layers grown on MgO, where the triangular-like switching is at the highest current density and less likely to occur.

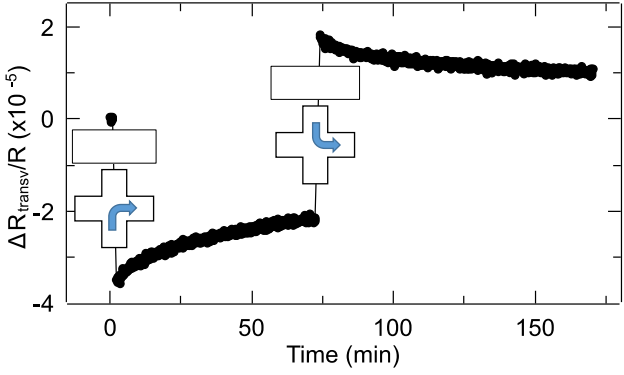

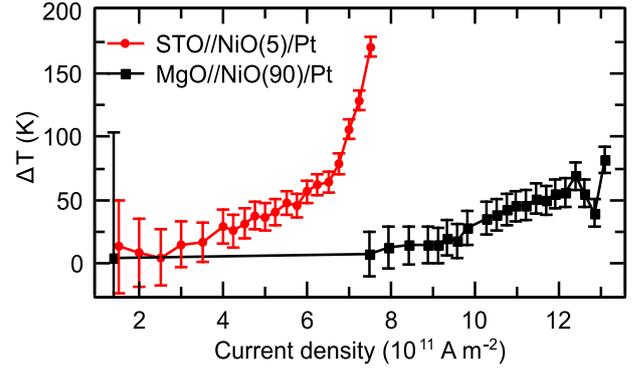

*Figure S9: Stability versus time of the switched states in a MgO//NiO(5)/Pt(2) sample, after the*

*Figure S10: Temperature increase during a single 1 ms pulse, performed by time-resolved*

**Stability of the switched states**
The stability of the switched states was tested after the application of 5 pulses at +45° and 5 pulses applied in the -45° direction in a MgO//NiO(5)/Pt(2) sample patterned with a Hall cross 8 µm wide, as shown in Fig. S9. The square pulses comprised a current density equal to $1.1 \times 10^{12}$ A m$^{-2}$, pulse width 1 ms, and time separation between two pulses in the pulse train equal to 10 s, to allow the sample to cool down and avoid thermal effects. The transverse resistance measurements were acquired approximately every 7 s. As one can see from Fig. S9, a slow relaxation process occurs on the time scale of hours. However, clearly distinguishable transverse resistance states are found even days after the application of pulses of different orientation.

**Current density threshold relation to heating effects**
In order to understand why different substrates show different current density thresholds for the triangular-like switching, we monitored the device resistance during the pulse application by means of an Agilent DSO6014A oscilloscope for a NiO(5)/Pt(2) sample grown on STO and patterned with a Hall cross 10 µm wide, and a NiO(90)/Pt(2) sample grown on MgO and patterned with a Hall cross 11 µm wide. First, we estimated the slope of the R vs T curve of the patterned devices, by measuring the resistance in a cryogenic environment in the temperature range 200 K – 300 K. Note that the resistance of the Pt is linear as a function of temperature for a wide temperature range around room temperature as confirmed by our measurements in the cryostat, so that the temperature variation is related to the relative resistance variation using the equation $\Delta T = k\Delta R/R$, even above the studied temperature range. After this R vs T calibration, we moved the sample to ambient temperature and we applied several series of 1 ms-long pulses with alternated current orientation every 5 pulses (±45°), increasing the current density every 20 pulses until the breakdown of the devices. By using a Keithley 6221 current source, allowing to set the pulse current to a desired value, the voltage measured by the oscilloscope is proportional to the resistance. By looking at the voltage at the pulse start and voltage at the pulse end, we were able to estimate the average temperature increase during the pulses, via the equation $\Delta T = k(V_{end} - V_{start})/V_{start}$. A voltage reducer made by two 1 MΩ resistor in series was connected in parallel to the sample in the case of the MgO sample, since the voltage exceeded the oscilloscope dynamic limit of 40 V, while connecting the oscilloscope in parallel to one of the two resistors. The results of this study are shown in Fig. S10. The heating of the samples grown on MgO(001) and STO(001) substrates is very different, with a much lower temperature increase when the samples are grown on MgO substrates, as resulting from the higher thermal conductivity of MgO, equal to 57 W m$^{-1}$ K$^{-1}$ compared to STO, whose conductivity is 11 W m$^{-1}$ K$^{-1}$. [6] This is reflected in the different current threshold densities of the switching (about 5x10$^{11}$ A m$^{-2}$ for the sample grown on STO and 9x10$^{11}$ A m$^{-2}$ for the sample grown on MgO), both occurring approximately when the temperature increase is equal to 35 ± 10 K. For both samples the breakdown of the devices occurred after the last point shown, i.e. at higher current density on MgO samples. This additionally confirms the lower heating experienced by samples grown on MgO substrates compared to samples grown on the STO for identical current densities. Note that also the thermal conductance of the thin NiO layer might have an effect (minor) on the temperature increase. The thermal conductivity of the NiO (~30 W m$^{-1}$ K$^{-1}$ at room temperature) [9] is intermediate between those of MgO and STO substrates.

**Domain structure of MgO(001)//NiO(001) thin films**

In Fig. 1a of the main text the x-ray absorption spectrum of a NiO layer 25 nm thick is shown, where the absorption at the Ni L$_2$ edge is different for linear horizontal (electric field parallel to the plane of the sample) and linear vertical (electric field inclined by 16° with respect to the normal of the plane of the sample) x-ray polarizations, because of the x-ray magnetic linear dichroism (XMLD) effect in the NiO film. We selected the energies for the XMLD – photoemission electron microscopy (PEEM) images by finding the maximum and minimum of the XMLD asymmetry spectrum, calculated as

$$XMLD_{spectrum} = \frac{I(LH) - I(LV)}{I(LH) + I(LV)}$$

Where I is the photoemission intensity at the detector. The maximum and minimum of the XMLD calculated with the formula above, are found at the energies $hv_1 = 869.4\ eV$ and $hv_2 = 871.0\ eV$, respectively. In Fig. S11a,b we show the direct intensity image, representing the morphology of the surface of the sample, together with a line profile of a particle on top of the surface. From the 10%-90% increase of the signal in the line profile we infer a resolution for this image better than 90 nm. To generate the XMLD images with better signal and information on how the antiferromagnetic domains are oriented, we acquired images with a single polarization (linear horizontal in the case of Fig. 4 of the main text) and two energies. The images were normalized and drift-corrected and the XMLD contrast was calculated according to the formula

$$XMLD_{images} = \frac{I(hv_1) - I(hv_2)}{I(hv_1) + I(hv_2)}$$

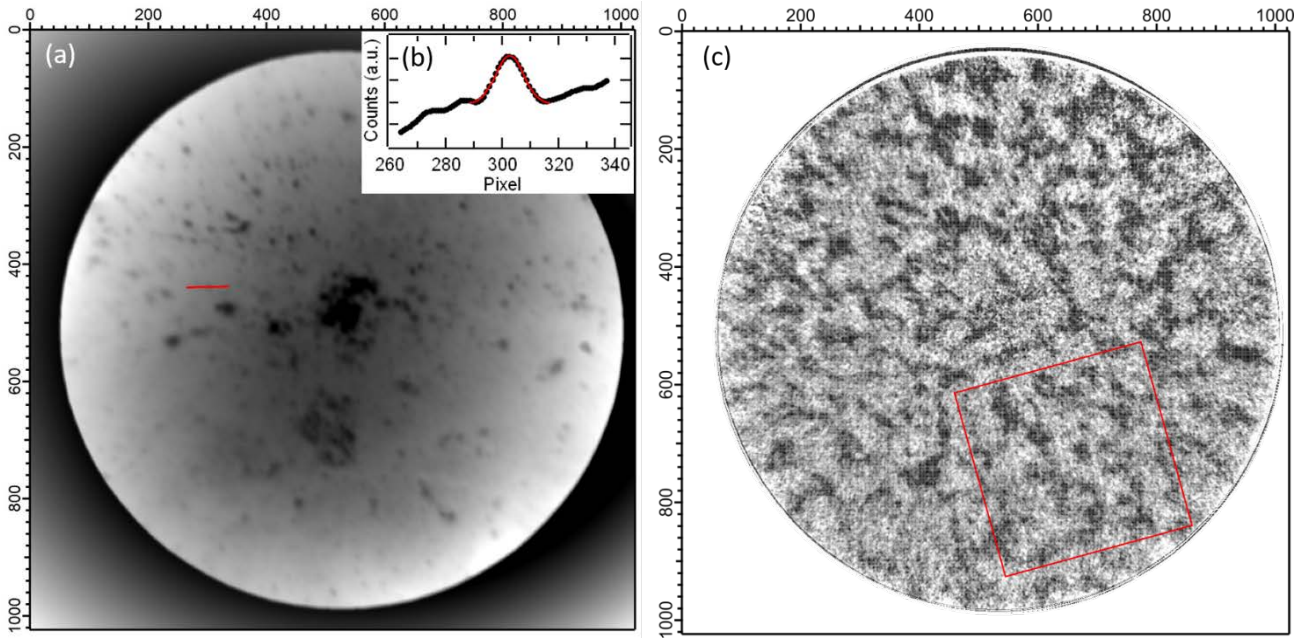

*Figure S11: (a) Direct intensity image used for the XMLD-PEEM measurements, in a field of view of 10 micron. The red line indicates the region where a line profile was taken. (b) Line profile of a particle on the surface,*

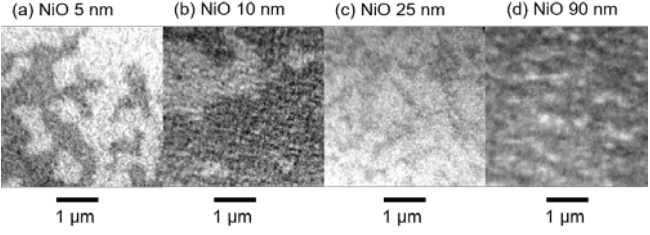

*Figure S12: Antiferromagnetic domain structure of NiO(001) films of different thicknesses, imaged by*

The result of this procedure for light with linear horizontal polarization is shown in Fig. S11c. We observe both XMLD contrast with linear horizontal and linear vertical polarizations, signaling that the orientation of the Néel vector in the NiO(001) domains grown on MgO is not only out-of-plane, as previously considered, [5] but more complex, possibly similar to the bulk structure which comprises 12 possible domains. We show in Fig. S12 the different antiferromagnetic domain structures observed by the XMLD-PEEM in all different NiO(001)/Pt thin film samples grown on MgO(001), where the NiO layer is 5, 10, 25 and 90 nm thick. The samples were grown together with the ones used for electrical measurements in Fig. 2 of the main text. The domain size does not have a simple monotonic dependence on the NiO thickness and we find large domains also in samples comprising thin NiO layers. Note that the presence of displaceable domain walls is evident in all samples, which is a prerequisite of our model.

**Theoretical model**

In this Section we derive the effective equation of motion for the domain walls in the presence of a spin-polarized current. We consider a compensated collinear antiferromagnet with two magnetic sublattices $\mathbf{M}_1$ and $\mathbf{M}_2$, whose magnetic state is fully described by the Néel vector $\mathbf{n} = \mathbf{M}_1 - \mathbf{M}_2$ ($|\mathbf{M}_1| = |\mathbf{M}_2| = M_s/2$). We assume that the magnetic anisotropy of the antiferromagnet is of an easy-plane type. In the absence of an applied current, an equilibrium state is represented by the equivalent domains with noncollinear Néel vectors $\mathbf{n}_A$, $\mathbf{n}_B$, etc (see Fig. 3a of the main text). We further assume that the domain walls separating domains A and B are of the Néel type (i.e. the Néel vector inside the domain wall rotates within the plane $\{\mathbf{n}_A, \mathbf{n}_B\}$).

The spin current is induced via the spin-Hall effect in the Pt electrode. We characterize the spin current with the vector of spin polarization $\mathbf{s}$ ($|\mathbf{s}| = 1$) and constant $H_{\text{curr}}$ which are related with the current density $\mathbf{j}$ as follows:

$$H_{\text{curr}}\mathbf{s} = \frac{\hbar \varepsilon \theta_H}{2 e d_{\text{AF}} M_s} \mathbf{j} \times \hat{z}, \qquad (1)$$

where $\hbar$ is the Planck constant, $d_{\text{AF}}$ is the thickness of the active layer of antiferromagnet, $0 < \varepsilon \leq 1$ is the spin-polarization efficiency, $\theta_H$ is the bulk spin Hall angle, $e$ is the electron charge. The axis $\hat{z}$ is directed along the film normal and, in the general case, it is different from the magnetic hard axis $\hat{Z}$. Note that, by using Eq. 1, we can calculate the effective staggered field for the current density we are using ($10^{12}$ A/m$^2$), considering 100% polarization efficiency (e.g., all spin-polarized electrons transfer their moments to NiO). For this and the following calculations we assumed that $d_{\text{AF}}$ =3.5 nm, $\theta_H$ =0.03, [10] spin polarization efficiency $\varepsilon$=1, $M_s = 5 \cdot 10^5$ A/m. We obtain a value below 5 mT, which is much smaller than the critical field (100 mT) required for coherent rotation, obtained from the anisotropy field values. [11,12]

The dynamics of the Néel vector in the presence of a spin current is described by the equation: [13,14]

$$\mathbf{n} \times [\ddot{\mathbf{n}} + \gamma \alpha_G H_{\text{ex}} \dot{\mathbf{n}} - c^2 \Delta \mathbf{n} - \gamma^2 H_{\text{ex}} M_s \mathbf{H}_\mathbf{n}] = \gamma^2 H_{\text{ex}} H_{\text{curr}} \mathbf{n} \times (\mathbf{s} \times \mathbf{n}) \qquad (2)$$

where $\gamma$ is the gyromagnetic ratio, $\alpha_G$ is the Gilbert damping constant, $H_{\text{ex}}$ is the intersublattice exchange field that keeps the vectors $\mathbf{M}_1$ and $\mathbf{M}_2$ antiparallel, $c$ is the velocity of the long-wavelength magnons. The effective field $\mathbf{H}_n \equiv -\partial U_{\text{ma}}/\partial \mathbf{n}$, where $U_{\text{ma}}$ is the energy density of the magnetocrystalline anisotropy, whose explicit expression depends on the symmetry. We additionally used the parameters exchange field H$_{\text{ex}}$ = 978 T, Gilbert damping $\alpha_G$ = 2x10$^{-4}$. [11,12,15]

*Current-induced contribution into the effective magnetic anisotropy*

We start from the analysis of the equilibrium homogeneous states in the presence of a current. From Eq. (2) it follows that the equilibrium orientation of the Néel vector is defined by the competition of two torques, the current-induced torque $\mathbf{T}_{\text{curr}} = M_s H_{\text{curr}} \mathbf{n} \times (\mathbf{s} \times \mathbf{n})$, and the torque $\mathbf{T}_{\text{an}} = M_s \mathbf{n} \times \mathbf{H}_\mathbf{n}$ created by the magnetic anisotropy field (see also Fig. 3a,b of the main text). The spin current can thus induce a static rotation of the Néel vector from the initial state $\mathbf{n}^{(0)}$ (in the absence of current) to a new state $\mathbf{n}^{(0)} + \Delta \mathbf{n}$. Such a rotation is reversible and can be associated with a potential energy of rotation. To find an explicit expression for the potential energy we calculate the virtual work $\delta W_{\text{curr}}$ produced by the current-induced torque $\mathbf{T}_{\text{curr}}$, $\delta W_{\text{curr}} = \mathbf{T}_{\text{curr}} \cdot \mathbf{e} \delta \theta$, in a process of quasistatic rotation around the $\mathbf{e}$ axis by a small angle $\delta \theta$. The infinitesimal variation of the Néel vector, $\delta \mathbf{n}$, is related with the rotation angle as follows: $\delta \mathbf{n} =$

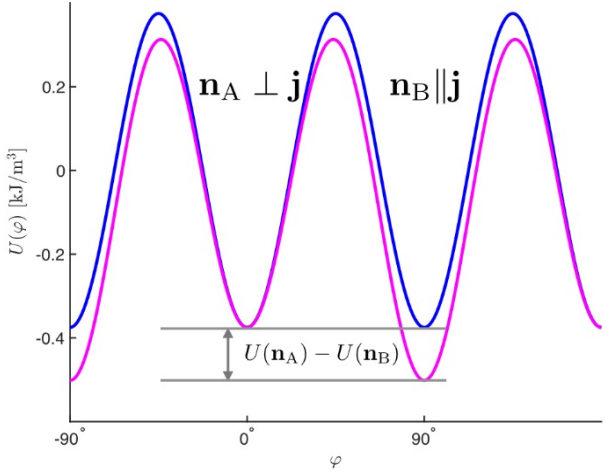

*Figure S13: Effective magnetic anisotropy energy calculated from Eq. (7) in the absence of current (blue line) and for current density* $j = 10^{12}$ *A/m²*

$\delta\theta \mathbf{e} \times \mathbf{n}^{(0)}$. Noticing that $\mathbf{n}^{(0)} \times \delta\mathbf{n} = \delta\theta[M_s^2 \mathbf{e} - (\mathbf{e} \cdot \mathbf{n}^{(0)})\mathbf{n}^{(0)}]$ and using the explicit expression for the torque we obtain:

$$\delta W_{\text{curr}} = \frac{H_{\text{curr}}}{M_s} \delta\mathbf{n} \cdot (\mathbf{s} \times \mathbf{n}^{(0)}). \quad (3)$$

Using further Eq. (1) we ultimately arrive at the expression for the current-induced potential energy density:

$$U_{\text{curr}} \equiv -\int_{\mathbf{n}^{(0)}}^{\mathbf{n}^{(0)}+\Delta\mathbf{n}} \delta W_{\text{curr}} = \frac{\hbar\varepsilon\theta_H}{2ed_{\text{AF}}M_s^2}(\hat{z} \times \mathbf{j}) \cdot (\mathbf{n}^{(0)} \times \Delta\mathbf{n}), \quad (4)$$

given in the main text of the paper. Note, that according to the definition, $U_{\text{curr}}$ is calculated with respect to a fixed initial state $\mathbf{n}^{(0)}$, which can be different in different domains.

*Current-induced energy splitting and ponderomotive force*

In this section we calculate the splitting of the effective magnetic anisotropy energy $U = U_{\text{ma}} + U_{\text{curr}}$ for two different equilibrium orientations of the Néel vector, $\mathbf{n}_A$ and $\mathbf{n}_B$, corresponding to two equivalent domains (see Fig. 3 in the main text). The magnetocrystalline anisotropy per unit volume of an easy-plane antiferromagnet is modeled as:

$$U_{\text{an}} = \frac{1}{2M_s} H_{\parallel} n_Z^2 - M_s H_{\perp} f(n_X, n_Y), \quad (5)$$

where the constants $H_{\parallel}$ and $H_{\perp}$ describe the out-of-plane and in-plane anisotropy fields, and the axis $\hat{Z}$ is perpendicular to the easy plane. For the calculations we used an out-of-plane anisotropy field $H_{2\parallel} = 0.16$ T and an in-plane anisotropy field $H_{\perp} = 0.012$ T. [11,12] The explicit expression for the dimensionless function $f(n_X, n_Y)$ depends on the symmetry of the antiferromagnet. For the analytical calculations we assume that $H_{\parallel} \gg H_{\perp}, H_{\text{curr}}$.

The spin current polarization $\mathbf{s}$ can be decomposed into the out-of-plane, $s_Z \hat{Z}$, and in-plane, $\mathbf{s}_{\perp}$, components. The out-of-plane component of spin current induces rotation of the Néel vector within the easy plane. In this case, as follows from symmetry considerations, both vectors $\mathbf{n}_A$ and $\mathbf{n}_B$ are rotated by the same angle. This component thus does not contribute into the energy difference $U(\mathbf{n}_A) - U(\mathbf{n}_B)$.

The in-plane component of the spin current induces out-of-plane rotation of the Néel vector $\Delta\mathbf{n} \approx n_Z \hat{Z}$ (taking into account that $H_{\parallel} \gg H_{\perp}, H_{\text{curr}}$). Substituting this expression into Eq. (5) we rewrite the current-induced contribution as

$$U_{\text{curr}} \approx \frac{H_{\text{curr}}}{M_s} n_Z \mathbf{n} \cdot (\hat{Z} \times \mathbf{s}_{\perp}), \quad (6)$$

Minimization of the effective magnetic anisotropy energy $U(\mathbf{n})$ with respect to the $n_Z$ component gives $n_Z \approx H_{\text{curr}} \hat{Z} \cdot (\mathbf{s}_{\perp} \times \mathbf{n}_{A,B})/H_{2\parallel}$. By substituting this expression into Eqs. (5) and (6), we can get the expression for the effective energy as a function of the in-plane component of the Néel vector:

$$U(\varphi) = -M_s H_{\perp} f(\varphi) - \frac{H_{\text{curr}}^2}{2H_{\parallel} M_s} \sin^2\varphi, \quad (7)$$

where $\varphi$ parametrizes in-plane component of the Néel vector, $n_X \approx \cos\varphi$, $n_Y \approx \sin\varphi$ and we assumed that $\mathbf{s}_{\perp} || \hat{X}$.

Eq. (6) shows that the effect of the current is analogous to the effect of the field which induces an additional uniaxial anisotropy within the easy plane. The angular dependence of the energy $U(\varphi)$ calculated for the tetragonal easy plane antiferromagnet with $f(n_X, n_Y) = (n_X^4 + n_Y^4)/M_s^4$ is shown in Figure S13.

In the general case, the difference between the two energy minima corresponding to A and B domains is described by expression:

$$U(\mathbf{n}_A) - U(\mathbf{n}_B) = \frac{H_{\text{curr}}^2}{2H_{2\parallel} M_s |\mathbf{j}_{\perp}|^2} [(\mathbf{j}_{\perp} \cdot \mathbf{n}_B)^2 - (\mathbf{j}_{\perp} \cdot \mathbf{n}_A)^2] \quad (8)$$

According to Eq. (8), the favorable domain is the one with the larger projection of the Néel vector on the current direction.

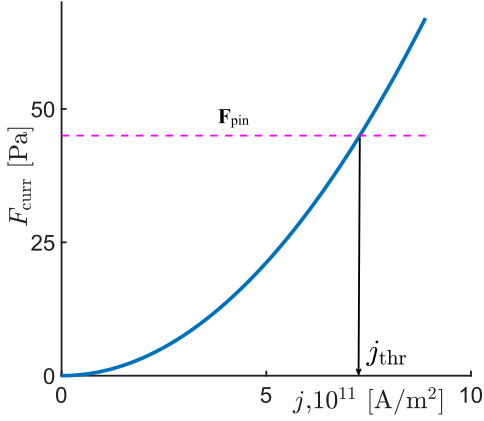

*Figure S14: Resulting force acting on the domain walls for **s** parallel to the easy plane ($s_Z=0$). The*

*Spin-current induced forces acting on the domain wall*
In this section we derive the explicit expressions for the forces acting on a single domain wall. We assume that the domain wall is flat, of the Néel type, and we calculate the force per unit area. Following the collective coordinate approach (for the details see Supplementary Materials in Ref. [16]) we introduce the linear momentum of the texture,

$$P_j = -\frac{1}{\gamma^2 M_s H_{ex}} \int \dot{\mathbf{n}} \cdot \partial_j \mathbf{n}\, d\xi \qquad (9)$$

which is conserved in the absence of a current. Here $\xi$ is the coordinate in the direction of the domain wall normal. The equation of motion for **P** is obtained from Eq. (2) by multiplying by $\mathbf{n} \times \partial_\xi \mathbf{n}/(\gamma^2 M_s H_{ex})$ and integrating along $\xi$:

$$\frac{dP_\xi}{dt} = -\alpha_G H_{ex} P_\xi + F_{curr}, \qquad (10)$$

where:

$$F_{curr} = \frac{H_{curr}}{M_s}\int_B^A \mathbf{s}\cdot \mathbf{n}\times \partial_\xi \mathbf{n}\, d\xi + U_{ma}(\mathbf{n}_A) - U_{ma}(\mathbf{n}_B) \qquad (11)$$

Decomposing **s** into in-plane $\mathbf{s}_\perp$ and out-of-plane $s_Z \hat{Z}$ components and using Eqs. (5, 8) and the considerations above, we represent the current-induced force $F_{curr} = F_{pond} + F_{DW}$ as a sum of the ponderomotive force $F_{pond} \equiv U(\mathbf{n}_A) - U(\mathbf{n}_B)$ and the chirality-dependent force:

$$F_{DW} = \frac{\hbar \varepsilon \theta_H}{2e d_{AF} M_s^2} \int_{\mathbf{n}_B}^{\mathbf{n}_A}(\hat{z}\times \mathbf{j})\cdot \mathbf{n}\times \partial_\xi \mathbf{n}\, d\xi \qquad (12)$$

In Eq. (12), which coincides with Eq. (1) of the main text, we express $H_{curr}$ and **s** in terms of material-related parameters. The current dependency of the resulting forces $F_{pond} \pm F_{DW}$ calculated for the domain walls with opposite chiralities and different $s_Z$ values are shown in Figs. S14 and S15.

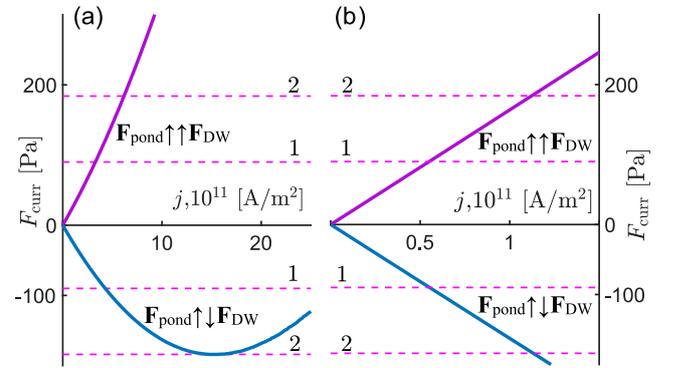

*Figure S15: Resulting force acting on the domain walls with opposite chiralities for $s_Z=0.087$ (a) and*